\documentclass[10pt]{mpi08} \usepackage{graphicx} 
\usepackage{cite,./mcite}  
\begin{document} \title{Jet reconstruction in heavy ion collisions (emphasis on Underlying Event background subtraction)} 
\author{M. Estienne$^1$}
\institute{$^1$SUBATECH, Ecole des mines, Universit\'e de Nantes, CNRS/IN2P3\\ 4 rue Alfred Kastler - BP 20722, 44307
       Nantes Cedex 3, FRANCE}
\maketitle
\begin{abstract}
A modification of the internal structure of jets is expected due to
the production of a dense QCD medium, the Quark Gluon Plasma, in
heavy-ion collisions. We discuss some aspects of jet reconstruction in
$p+p$ and $A+A$ collisions and emphasize the dramatically increased
contribution of the underlying event in nucleus-nucleus collisions as
compared with the vacuum case. We conclude with its consequences on
the full jet spectrum and fragmentation function extraction at LHC.
\end{abstract}
\section{Motivations for jet studies}
\subsection{The phenomenon of jet energy loss in heavy-ion collisions}
Non-perturbative lattice QCD calculations indicate that a deconfined
state of matter, the Quark Gluon Plasma (QGP), may exist at very high
temperatures and energy densities. This state of matter is expected to
be formed in the heart of an ultra-relativistic heavy-ion collision,
when the energy density is the largest.  Since 2000, the Relativistic
Heavy-Ion Collider (RHIC) has collected impressive results, which has
led to the discovery of a new state-of-matter of very small
viscosity~\cite{ref:white}. Among the observables which have led to
such a conclusion, the jet quenching effect is one of the most
relevant as it has highlighted the production of a dense medium in
interaction. One of the first computations of the radiative energy
loss of high-energy quarks in a dense medium was proposed by Gyulassy
et al.~\cite{ref:gyu,ref:wan} in the early nineties. Since then many
approaches have been developed to determine the gluon radiation
spectrum of a hard parton undergoing multiple
scattering~\cite{ref:BDMPS,ref:zak,ref:wid,ref:glv}. The experimental
consequence of these processes is a significant suppression of large
transverse momentum ($p_{T}$) hadrons in heavy-ion collisions (HIC)
highlighted through the measurement of the nuclear modification factor
or two and three particle correlations~\cite{ref:rcp,ref:azi}. Even
though we can nowadays claim that a dense medium has indeed been
produced and somehow characterized, a plethora of questions remains:
does energy loss result from few strong scatterings in the medium or
multiple soft ones ? How does it depend on the medium-length ? What is
the energy loss probability distribution of the partons ? They
motivate the necessity to call for some more discriminating, and
differential observables to characterize the QGP.

Moreover, the ``leading particle'' physics which has been studied at
RHIC until 2008 presents some limitations known as {\it surface} and
{\it trigger} biases~\cite{ref:surf,ref:trig}. Ideally, the analysis
of reconstructed jets on an event by event basis should increase the
sensitivity to medium parameters by reducing the trigger bias and
improve our knowledge of the original parton 4-momentum.

\subsection{Jets in a heavy-ion collision and the Underlying Event background}

In QCD, jets are defined as cascades of partons emitted from an
initial hard scattering followed by fragmentation. In HIC, parton
fragmentation is modified relative to the vacuum, due to the presence
of the hot QCD medium. After the overlap of the two incoming nuclei,
the quarks and gluons produced in the initial nucleon-nucleon ($N+N$)
hard scatterings propagate through the dense color field generated by
the soft part of the event. Consequently, the medium should affect
the fragmentation process of hard partons and has drastic effects on
the jet structure itself. (i) A softening of the fragmentation
function is expected leading to the suppression of production of high
$p_{T}$ particles as well as a numerous production of soft
particles. A first attempt to model medium-modification fragmentation
processes by Borghini \& Wiedemann was the determination of the single
inclusive hadron spectrum inside jet - known as Hump-Backed Plateau
(HBP) - in HIC~\cite{ref:hbp}.  This aspect will be addressed in
section~\ref{par:FF} at the level of the experiment. (ii) A jet
broadening (inducing out-of-cone radiations) is expected as one should
observe a redistribution of the particles inside the jet relatively to
its axis. A modification of the transverse shape of the jet ($k_{T}$
spectrum) or its particle angular distribution can be
studied~\cite{ref:yr}. (iii) In case of sufficiently strong energy
loss scenarii, it could have consequences on the jet reconstruction
itself and reduce the expected jet rate. (iv) As di-jet pairs have
different path lengths in medium and as energy loss is a stochastic
process, the di-jet energy imbalance should be increased and
acoplanarity induced.

Ideally, a direct measurement of these modifications should be
possible.
However, the picture is more complicated due to the presence of the
soft Underlying Event (UE). The UE and its fluctuations will induce
important bias on the jet identification. It will be extensively
discussed in section~\ref{par:UE}. The expected jet reconstruction
performances in $p+p$ in the ALICE experiment are first discussed in
section~\ref{par:pp}. Note that the jet energy-scale, one of the main
sources of uncertainty in any jet spectrum measurement will not be
discussed here. ATLAS and CMS results will not be commented
either. More information can be found elsewhere~\cite{ref:atlas}.

\section{Jet reconstruction performances with calorimetry}
\label{par:pp}
\subsection{Experimental apparatus and tools}
Full jet measurement in heavy-ion experiments has become possible very recently
thanks to the insertion of an electromagnetic calorimeter (EMC) in the
STAR experiment at RHIC~\cite{ref:pp,ref:AA}.  STAR has demonstrated
the feasibility of such measurement combining its charged particle
momentum information from its Time Projection Chamber (TPC) and the
neutral one from the EMC, publishing the first measurement of the
inclusive jet spectrum for the process $p+p$ (both polarized)
$\rightarrow$ jet + X at $\sqrt{s} = 200$~GeV with a $0.2$~pb$^{-1}$
integrated luminosity~\cite{ref:pp}. The spectrum of pure power law
shape is in agreement with NLO calculations (within the error bars).\\
As STAR, ALICE is a multipurpose heavy-ion experiment~\cite{ref:ali}.
Its central barrel mainly equipped of a large TPC and a silicium inner
tracking system covers a full azimuthal acceptance but is limited to
the midrapididity region ($|\eta| < 0.9$). It has a large $p_{T}$
coverage ($\sim$ 100~MeV/$c$ to $\sim$ 100~GeV/$c$) with a $\delta
p_{T} / p_{T}$ resolution of few percents (still below 6\% at
100~GeV/$c$)~\cite{ref:surf}. The capabilities of ALICE to disentangle
particles down to very low $p_{T}$, where strong modifications of the
fragmentation function are expected, should lead to a very precise
measurement of the number of particles inside a jet.  More recently,
the insertion of an electromagnetic calorimeter to collect part of the
neutral information and to improve the trigger capabilities of ALICE
has been accepted as an upgrade. The EMCal is a
Pb-scintillator sampling EMC ($|\eta| < 0.7, 80^{\circ}
< \phi < 190^{\circ}$) with a design energy resolution of $\Delta E/E
= 11\%/\sqrt{E}$ and a radiation length of $\sim$ $20~
X_{0}$~\cite{ref:tdr}. It contains $\sim$13k towers in Shashlik
geometry with a quite high granularity ($\Delta\eta \times \Delta\phi
= 0.014 \times 0.014$). The official ALICE jet finder is a UA1 based
cone algorithm which has been modified in order to include the neutral
information during the jet finding procedure. 
\subsection{Jet signal degradation and energy resolution in $p+p$ collisions}
\label{par:resopp}
\begin{figure}[h]
\vspace*{-0.7cm}
\hspace{-2mm}
\begin{center}
\includegraphics[scale=0.27, angle=0]{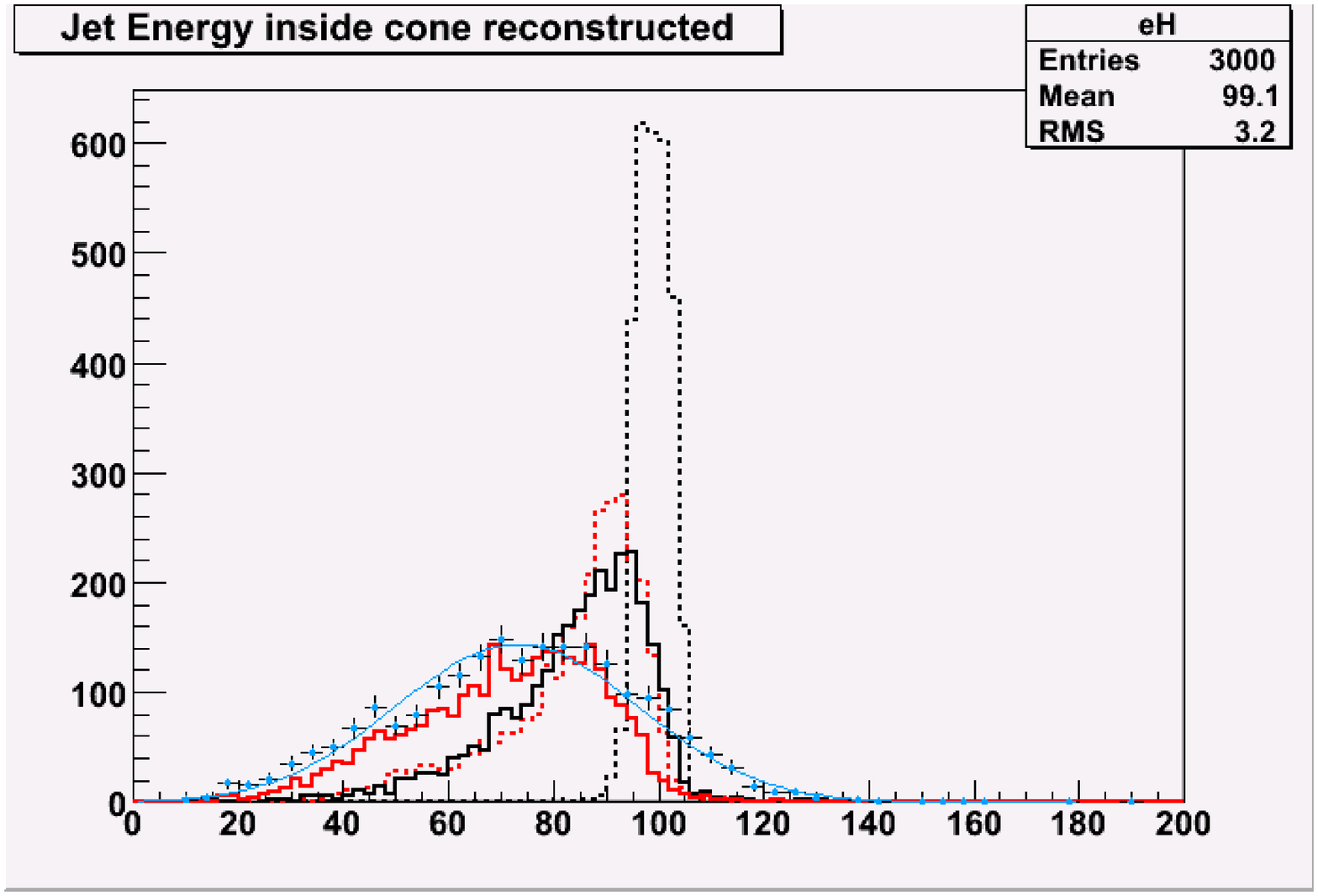}   
\includegraphics[scale=0.34, angle=0]{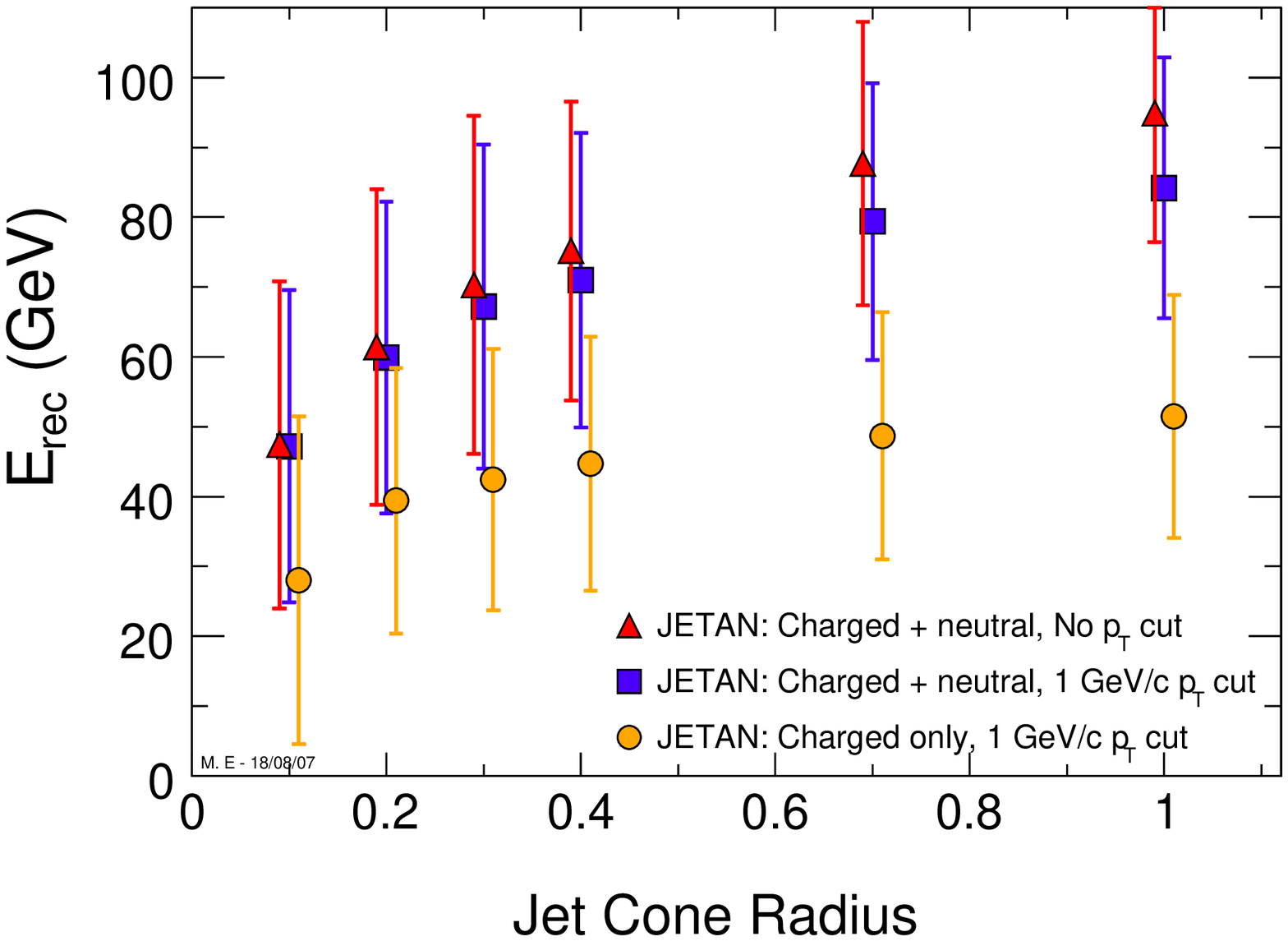}
\caption{\underline{Left}: cone energy of $100$~GeV jets reconstructed with {\small PYCELL} with $R = 1$ (dark dashed line), with the ALICE cone finder with detector inefficiencies and acceptance included in the simulation with $R = 1$ (red dashed), without detector effects but $R = 0.4$ (dark full), with both effects (red full). The markers shows the result from a full simulation. \underline{Right}: cone energy of $100 \pm 5~$GeV fully simulated jets vs R for the three cases described in the text.}
\label{fig:fluctuations}
\end{center}
\end{figure}
Jet reconstruction is highly influenced by the high multiplicity of an
event and by the charged-to-neutral fluctuations for jets in which the
neutral fraction (or part of it) can not be measured. Due to its
detector configuration, ALICE will be able to reconstruct two types of
jets. Using the charged particle momentum information, the production
of {\it charged jets} will be studied. As the charged particle plus
EMCal configuration is almost blind to neutrons and $K^{0}_{L}$, ALICE
will also measure {\it charged+neutral jets} but will miss part of the
neutral energy. In both cases and in elementary collisions, the
charged-to-neutral fluctuations which dominate will give rise to a low
energy tail in the reconstructed jet energy. Such effects should be
enlarged by limited detector acceptance and inefficiency and analysis
cuts which cause other types of fluctuations. To get a basic and
qualitative understanding of the signal fluctuations for jets
reconstructed in $p+p$ collisions at LHC, we have undertaken a fast
simulation of $100 \pm 5~$GeV jets using PYTHIA as event generator for
different cuts and detector configurations. Such features are
illustrated in Fig.~\ref{fig:fluctuations} (left) which shows the
distribution of the jet energy reconstructed in a cone of radius $R$
and compared with the result from a full detector simulation described
below.

Jets were first reconstructed with a simple jet finder available in
PYTHIA (PYCELL) with $R = 1$ using the momentum and energy information
from charged and neutral particles (neutrons and $K^{0}_{L}$ excluded)
(full black line). For the sample of simulated events which include
detector acceptance cuts and reconstructed track inefficiency (not
studied separately here), keeping R=1 for the jet reconstruction, one
or several of the leading jet particles are not reconstructed and do
not contribute to the cone energy. It leads to its broadening and a
low energy tail (red dashed curve). The use of a limited cone radius
during the jet finding procedure enhances collimated jets and also
leads to a low energy tail of the cone energy distribution (black
dashed line).  The full red curve shows the combination of all the
effects on the reconstructed jet energy keeping the jets which center
falls inside the EMCal acceptance. The reconstructed energy results in
an almost gaussian response function of resolution defined as $\Delta
E / E = r.m.s./<E>$ of $\sim$ 33\%. It can be improved selecting only
the jets fully contained in the EMCal as discussed below.
\begin{figure}[h] 
\vspace{-2mm}
\hspace{-2mm}
\includegraphics[scale=0.35, angle=0]{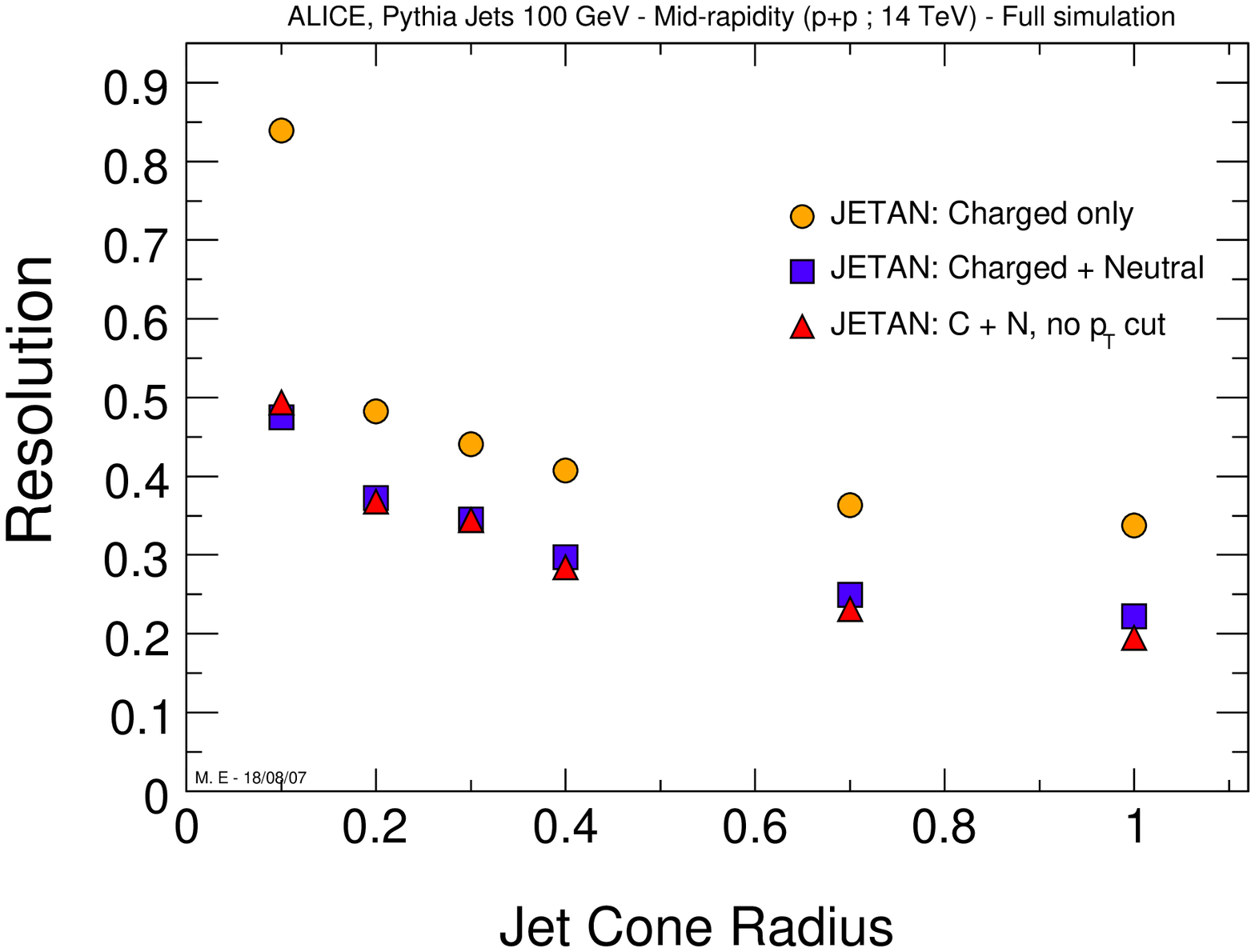}   
\includegraphics[scale=0.3,width=7cm,height=5.2cm, angle=0]{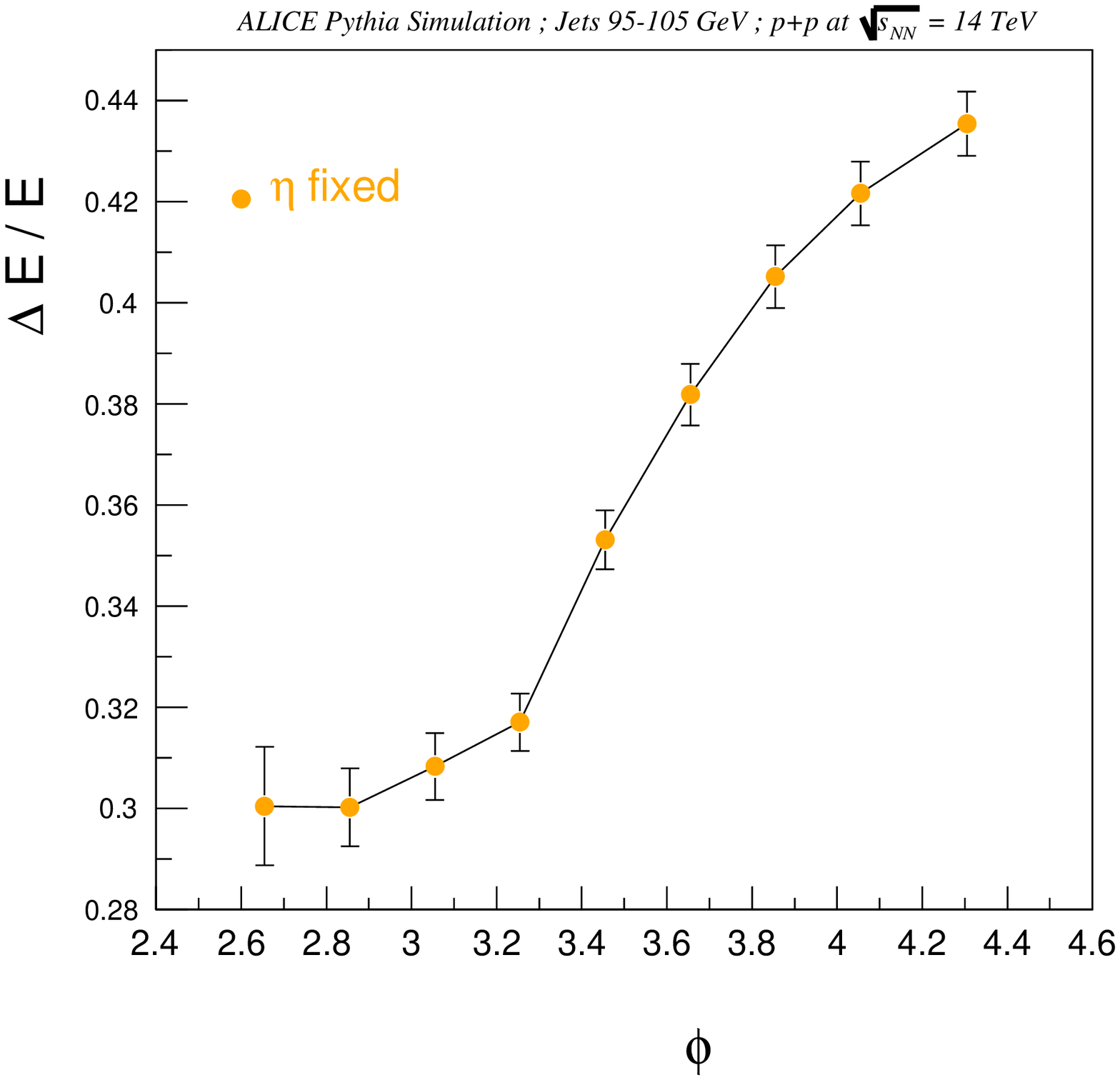}
\caption{\underline{Left}: jet energy resolution of $100~$GeV jets from a full ALICE simulation vs R for the three cases described in the text. \underline{Right}: jet energy resolution as a function of the accepted $\phi$ window of the center of the jet reconstructed.}
\label{fig:resolution}
\end{figure}

In the following, we present results obtained with a complete
simulation and reconstruction chain using PYTHIA as event generator
and GEANT3 for the detector responses for the generation of
monoenergetic jets of $50, 75$ and $100\pm 5~$GeV. The $\pm$5~GeV
uncertainty on the simulated jet energy will be implicit
below. Figure~\ref{fig:fluctuations} (right) presents the cone energy
reconstructed vs cone radius in three experimental conditions: with
charged particles only and 1 GeV/$c$ $p_{T}$ cut on their momentum
(circles), with charged plus EMCal configuration and 1 GeV/$c$ $p_{T}$
cut (squares) and with charged plus EMCal without $p_{T}$ cut. The
error bars are the r.m.s. of the energy
distributions. Figure~\ref{fig:resolution} (left) shows the same study
but for the resolution. As already discussed, reconstructing jets from
charged particles only enhances the number of jets with a larger than
average charged particle fraction. Increasing $R$ of course increases
the mean reconstructed energy and improves the resolution but one
reconstructs at best an energy below 50\% of the input energy. These
charged-to-neutral fluctuations lead to a resolution of $\sim 40\%$
for $R = 0.4$, improved to 30\% by the inclusion of neutral particles
in the jet finding procedure.  For $R = 1$, in the case charged +
neutral without $p_T$ cut, the resolution is at best of 20\% but part
of the neutral information is lost as the jet is not fully collected
within the calorimeter. The impact of the finite energy resolution on
the full reconstructed jet spectrum will be quickly discussed in
section~\ref{par:spectrum}.

The limited EMCal acceptance effect on the resolution of the
reconstructed jet energy has been studied
previously~\cite{ref:kiev}.
We have shown that as long as the jet center is taken inside the
EMCal, even if part of its energy is outside it, the resolution is
still close to 30\%. As long as the center of the jet can be taken
outside the EMCal acceptance, the resolution degrades and
asymptotically reaches the charged particles only case in the full TPC
acceptance (Fig.~\ref{fig:resolution} (right)).
\section{The underlying event in $A+A$ collisions}
\label{par:UE}

\subsection{The background in $A+A$ collisions}
Jet reconstruction in HI collisions is more complicated than
in elementary systems as the UE dramatically
changes. The reconstruction is dominated by the influence of
the high multiplicity.
A rough assessment of the energy of the UE inside R = 1 at RHIC based
on $dE_{T}/d\eta = 660$~GeV at mid-rapidity~\cite{ref:sps} gives
$E_{UE} = 1/(2\pi) \times \pi R^{2} \times dE_{T}/d\eta \sim$ 330~GeV.
A linear or logarithmic extrapolation of the charged particle rapidity
density from the available data at FOPI, SPS and RHIC~\cite{ref:sps}
allows to estimate an $E_{UE}$ between 500~GeV and 1.5~TeV at LHC. In
the extreme case, the UE is a 4-fold higher than at RHIC however the
growth of the cross-section for hard processes is more
dramatic. The substantial enhancement in the jet cross-section
significantly improves the kinematics reached for jet
measurement at LHC allowing the reconstruction of high-energy jets 
above the uncorrelated background on an event by event basis
with good statistics. 

Not only the multiplicity differs from $p+p$ collisions but the
physics phenomena. First, the simple fact that the impact parameter
varies event-by-event for a given centrality class implies some
fluctuations in the UE ($\propto R^{2}$). All the well known
correlations to the reaction plane and the azimuthal correlations between
two and three particles at momenta below 10 GeV/$c$ drag some structures
inside what can be denoted as background for our jet studies. They
are region-to-region fluctuations and are proportional to R. Moreover,
the main sources of region-to-region fluctuations are the Poissonian
fluctuations of uncorrelated particles also proportional to R. To
optimize the jet identification efficiency, the signal energy has to be
much larger than the background fluctuations $\Delta E_{bckg}$.
The energy of the UE and its fluctuations inside a given cone can be
considerably reduced by simply reducing R in the jet finding procedure
and applying a $1$ or 2~GeV/$c$ $p_{T}$ cut on charged
hadrons~\cite{ref:surf,ref:bly}.
However, they both imply some signal fluctuations whose effects have
been discussed above. The jet finding procedure in a HI environment is
thus essentially based on two steps. First, a $p_{T}$ cut and a
limited R are applied. Then, during the iteration procedure in the jet
finding algorithm which has been optimized accordingly, the remaining
energy of the UE outside the jet cone is estimated statistically or
event by event and is subtracted from the energy of the jet inside its
area at each iteration. Note that the use of a $p_{T}$ cut is
potentially dangerous for a quenching measurement~\cite{ref:AA} so
that new background subtraction technics based on jet areas should be
prefered and investigated to improve our measurement~\cite{ref:area}.
\subsection{Understand the background fluctuations}
\begin{figure}[h]
\begin{center}
\includegraphics[scale=0.6, angle=0]{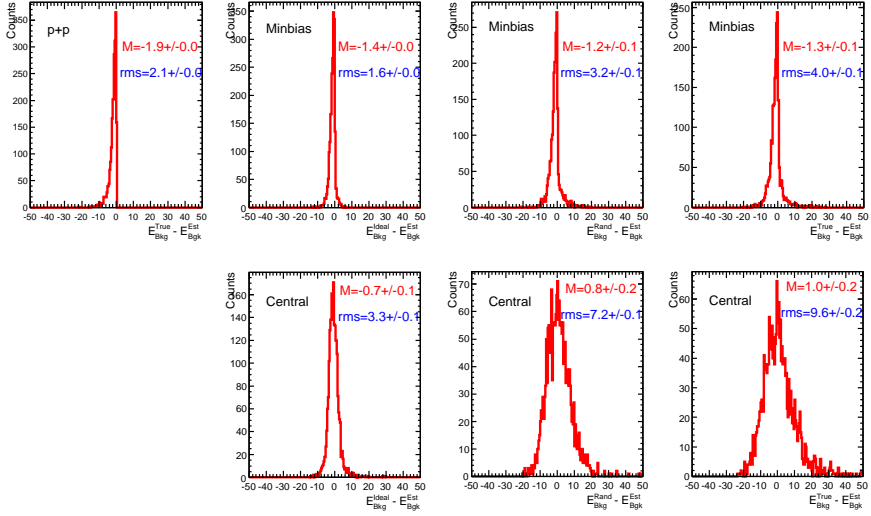}  
\caption{$E^{X}_{bgk} - E^{Est}_{bgk}$ for $p+p$, $Pb+Pb$ Minbias and Central collisions obtained from a full ALICE simulation. $E^{X}_{bgk}$ has been extracted in three X cases presented in the text.}
\label{fig:bckg}
\end{center}
\end{figure}
The validity of our background subtraction procedure applied in the
EMCal acceptance has been tested on three simulated data
sets~\cite{ref:in}. The full PYTHIA simulation of 100~GeV jets at
$\sqrt{s} = 14~$TeV has been used to mimic $p+p$
collisions. Similarly, we processed full Minbias and Central HIJING
simulations at $\sqrt{s_{NN}} = 5.5~$TeV to reproduce $Pb+Pb$ events
at LHC in the EMCal acceptance in which we embed PYTHIA events
to simulate the hard processes. The small change in the event
multiplity between $p+p$ and $Pb+Pb$ Minbias collisions does not
extensively increase the fluctuations in Minbias, unlike
Central compared with Minbias where a factor of $4 - 5$ in the
multiplicity is expected to drive an increase of a factor of $2 - 2.2$
in the fluctuations.

The later assertion has been tested and part of the obtained results
are presented in Fig.~\ref{fig:bckg}. We define the total fluctuations
as $\Delta E_{Tot} = \Delta E_{Sig} + \Delta E_{Bkg}$ (1).
One can estimate the variations of fluctuations between Minbias and
Central knowing the $p+p$ case. $\Delta E_{Bkg} = E_{Bkg}^{X} -
E_{Bkg}^{Est}$ has been estimated from three different methods $X$,
using an ($\eta,\phi$) grid filled with the HIJING particle
information output where the background energy inside a cone of radius
R is estimated by summing the energy (i) of all cells inside the grid
and scaling the total energy to the jet cone size ($X = Ideal$) ; (ii)
inside the cone taken randomly in the grid ($X = Rand$) ; (iii) inside
the cone centered on the jet axis (beforehand found by the jet finder)
($X = True$). The distributions are presented in the 6 right pannels
of Fig.~\ref{fig:bckg} for the $Ideal$ (left), $Rand$ (center) and
$True$ (right) cases respectively, and for Minbias (top) and Central
(bottom) collisions. The same exercise has been applied on a grid only
filled with $p+p$ events. The distribution of $\Delta E_{Bkg} =
E_{Bkg}^{True} - E_{Bkg}^{Est}$ is presented in the most left hand
panel. The mean value obtained for the distributions of Minbias data
are systematically negative. Clearly the jet algorithm over-estimates
the background compared with the three cases due to out-of-cone signal
fluctuations which does dominate as emphasized in the $p+p$
case. Going from the $Ideal$ to the $True$ case, the region-to-region
fluctuation effects increase the r.m.s. These fluctuations are less
pronounced in the $Ideal$ case which gives a mean value of the
background event by event. From Minbias to Central data, a factor of
$2 - 2.2$ in the r.m.s. is observed, as expected, validating our
background subtraction method. In Central, the fluctuations are thus
dominated by the event multiplicity. It is indeed observed in the mean
values which become positive with a large positive tail from the
$Ideal$ to the $True$ cases. In Central data, the background is thus
under-estimated by the jet algorithm so that the final cone energy is
over-estimated.
\subsection{Expected performances in $Pb+Pb$ collisions at LHC}
\label{par:performances}
\begin{figure}[h]
\begin{center}
\hspace{-4mm}
\includegraphics[scale=0.33, angle=0]{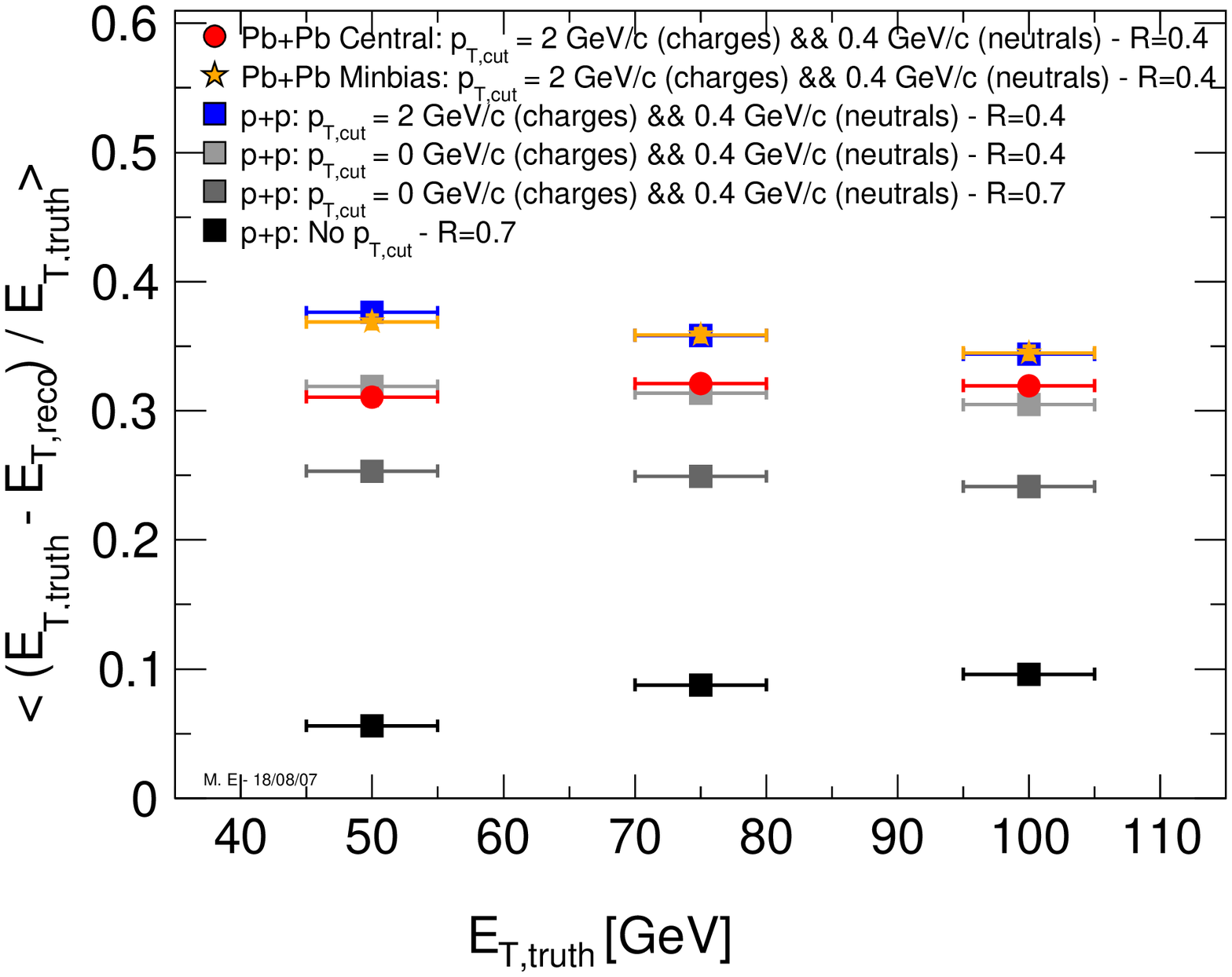} 
\includegraphics[scale=0.33, angle=0]{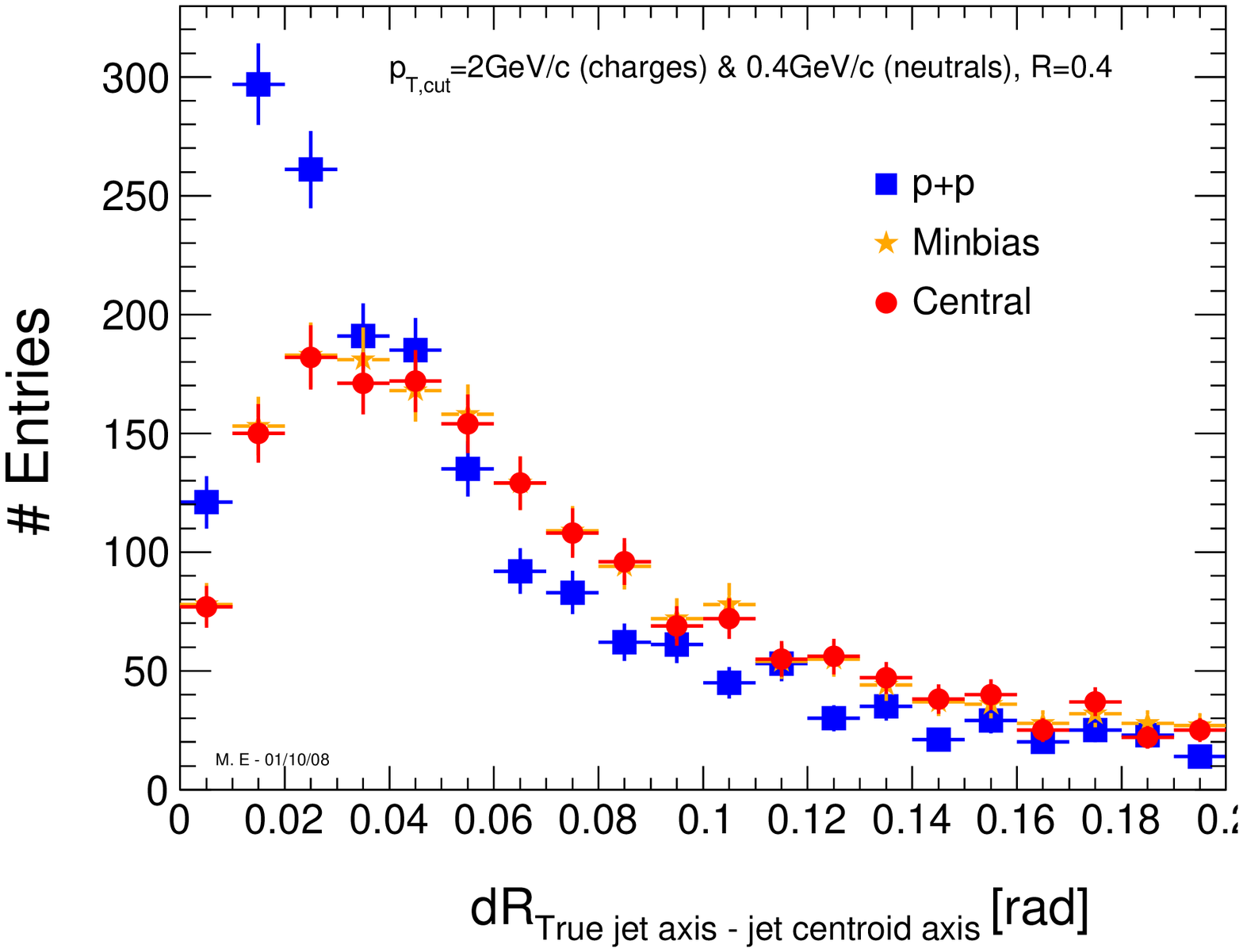} 
\caption{\underline{Left}: jet reconstruction efficiency as a function of $E_{T,truth}$ for the cases quoted in the top left legend of the figure. \underline{Right}: distance in $\eta$-$\phi$ space between the directions of the reconstructed jet axis and the true one in $p+p$ (squares), $Pb+Pb$ Minbias (stars) and $Pb+Pb$ Central (circles) collisions. }
\label{fig:efficiency}
\end{center}
\end{figure}
Figure~\ref{fig:efficiency} (left) presents what is defined as the
``jet reconstruction efficiency'' ($(E_{T,truth} - E_{T,reco}) /
E_{T,truth} = 1 - Efficiency$) as a function of the input jet energy,
$E_{T,truth}$, for the 3 input jet energies $50, 75$ and $100 \pm
5~$GeV. The Minbias and Central $Pb+Pb$ cases are compared with the
$p+p$ one for which a systematic study of the analysis cuts has also
been performed. Jets have been reconstructed using the ALICE UA1 cone
finder including both charged and neutral particles. The efficiency
obtained without $p_{T}$ cut and $R = 0.7$ (black squares) smoothly
increases when the input jet energy increases and reaches $10 \%$ for
100~GeV jets. It is enhanced by a factor of $3$ to $5$ after the
application of a $p_{T}$ cut of 0.4~GeV/$c$ on neutral particles (dark
grey squares). The reduction of $R$ to $0.4$ (light grey squares)
increases the efficiency (which becomes flat vs $E_{T,truth}$) to
$\sim 30 \%$ as less input jet energy is reconstructed. The efficiency
worsens moreover when a $p_{T}$ cut on the charged particles is
applied (blue squares) as part of the signal is again cut. In these
cases the reconstructed energy is under-estimated by the algorithm and
the out-of-cone fluctuations from the signal dominate. As expected in
Fig.~\ref{fig:bckg}, no significant discrepancies between $p+p$ and
$Pb+Pb$ Minbias data samples (stars) are observed whereas the
efficiency in Central (circles) is improved because the background
subtraction procedure over-estimates the cone energy and the
background fluctuations dominate. In Minbias, both effects compensate.

In order to understand how the fluctuations affect the jet
reconstruction, the distributions of the reconstructed jet axis minus
the input jet axis have been studied in the 6 previous cases. Both the
$p_{T}$ and radius cuts on $p+p$ data affects a bit
the jet reconstructed axis but the effect is
small. Figure~\ref{fig:efficiency} shows the distributions for
the Minbias and Central cases compared with the $p+p$ one. It clearly
shows that the reconstructed jet axis in both cases is biased. Using a
small radius, the jet algorithm maximizes the energy by shifting the
jet (centroid) axis.
\begin{figure}[h]
\begin{center}
\hspace{-2mm}
\includegraphics[scale=0.32, angle=0]{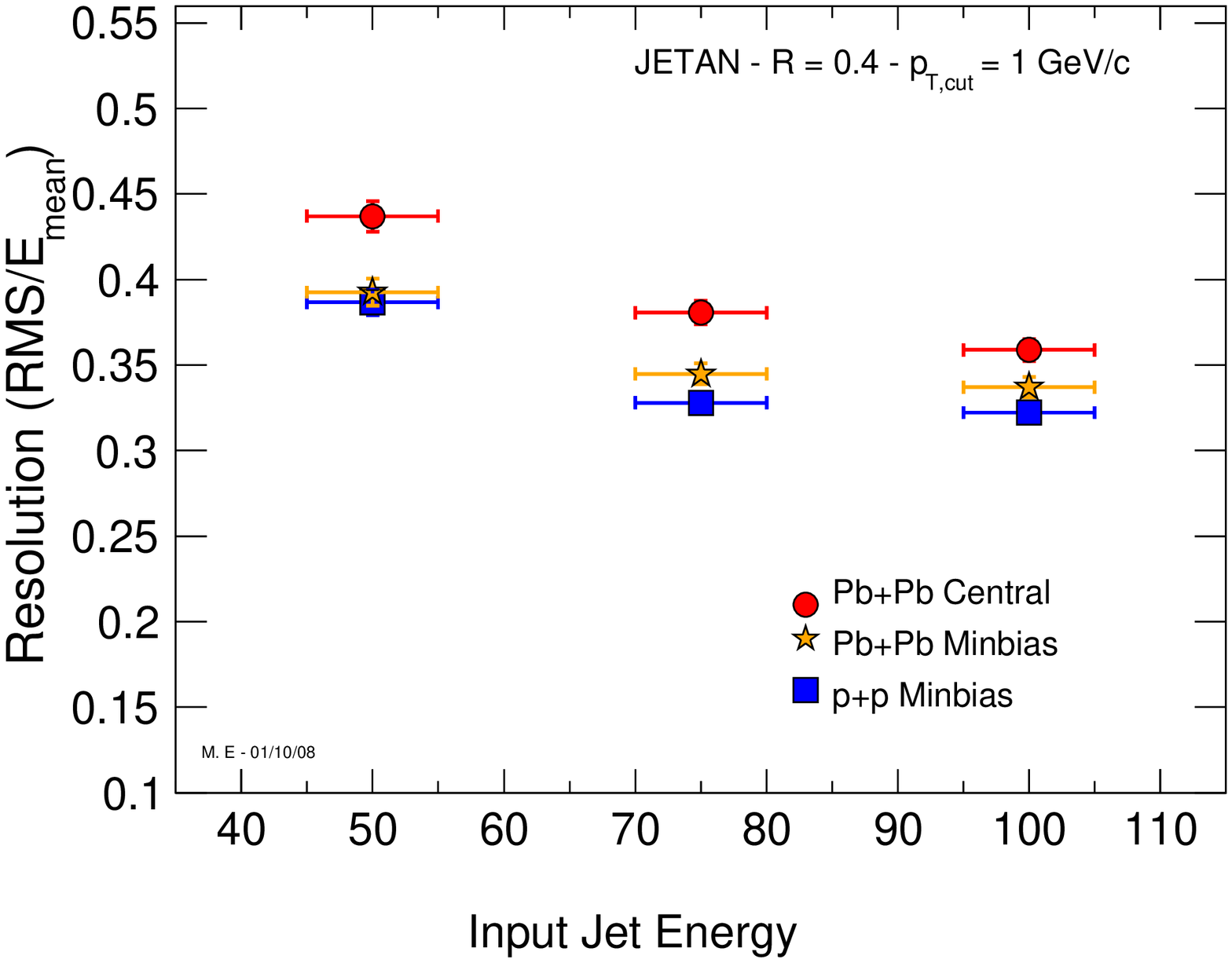}  
\includegraphics[scale=0.32, angle=0]{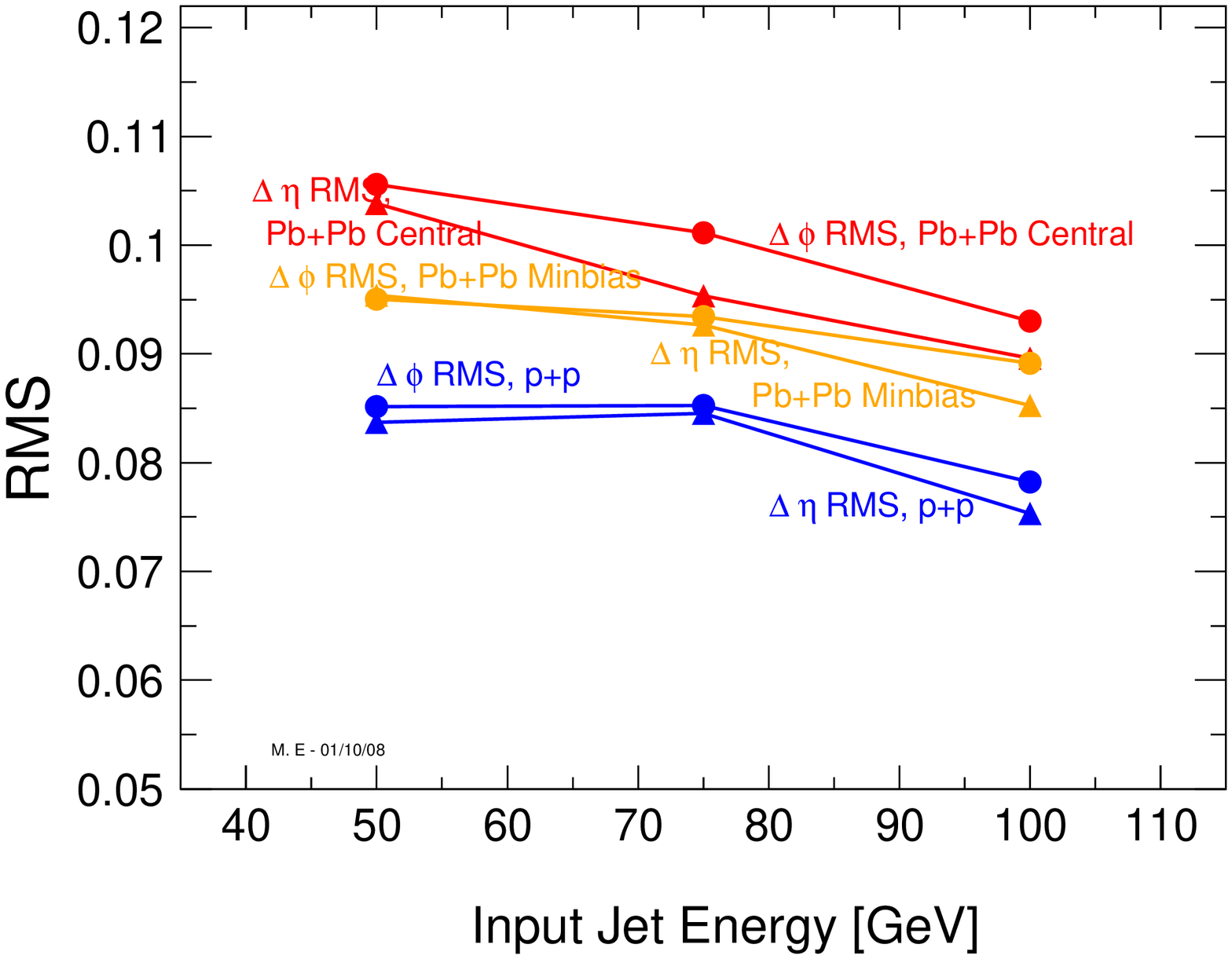}
\caption{\underline{Left}: jet energy resolution versus the input jet energy of $50, 75$ and $100 \pm 5~$GeV for $p+p$ (squares), $Pb+Pb$ Minbias (stars) and $Pb+Pb$ Central (cirles) collisions. 
\underline{Right}: resolutions in $\eta$ and $\phi$ of the jet direction.}
\label{fig:finalreso}
\end{center}
\end{figure}
In the different systems studied, the evolution of the expected jet
energy and angular resolutions versus $E_{T,truth}$ and the system
multiplicity are presented in Fig.~\ref{fig:finalreso} (left) and
(right). The jets have been reconstructed using a $p_{T}$ cut of
1~GeV/$c$ and $R = 0.4$. All the jets which centers lied inside the
EMCal acceptance were considered. The reconstructed energy resolution
worsens from $100~$GeV to $50~$GeV jets in the 3 systems. Contrary to
the jet reconstruction efficiency, the energy resolution degrades as
expected from $p+p$ to $Pb+Pb$ Central because of background
fluctuations.  For $100~$GeV jets, we obtain an energy resolution in
$p+p$ of $\Delta E_{p+p} \sim 32.5
\%$. The Minbias one allows to estimate the Central one to
$\Delta E_{Cent} \sim 35.8 \%$ using equation (1) in agreement with
the resolution of $36.4 \%$ obtained in Fig.~\ref{fig:finalreso} (left)
validating our background subtraction method.  
Figure~\ref{fig:finalreso} (right) presents the r.m.s. of the distributions $\Delta \eta =
\eta_{truth} - \eta_{reco}$ (triangle) and $\Delta \phi = \phi_{truth}
- \phi_{reco}$ (circle).
An accurate reconstruction of the jet direction in the three
systems is obtained though it is slightly deteriorated from p+p to Minbias and
Central. Indeed, the dominating background fluctuations maximize
the jet energy by shifting its reconstructed direction as observed in
Fig.~\ref{fig:finalreso}.
\section{Full jet spectrum and fragmentation function}
\label{par:FF}
\subsection{A smeared jet spectrum}
\label{par:spectrum}
The results presented so far do not take into account the jet cross
section distribution as 1/$p_{T}^{\alpha}$ with $\alpha \sim 5.7$ and beyond at
LHC.  We note that within a 1$\sigma$ fluctuation of the energy the
jet production cross section varies by almost
twofold~\cite{ref:surf}. Therefore, it is essential to take into
account the production spectrum to truly evaluate the meaningful jet
energy resolution and reconstruction efficiency. In particular, jets
in the low energy tail of the resolution function are buried below
lower energetic jets with much higher production cross section and,
hence, the amount of jets in these tails is a measure of the
reconstruction inefficiency.

In order to extract the jet production spectrum, 12 bins of
$p_{T-hard}$ from 40 to 220~GeV have been simulated with PYTHIA 6.2
CDF Tune A in the 2$\rightarrow$2 processes. The simulated data have
then been treated in the full detector chain of GEANT3 before
reconstruction using the official ALICE jet finder including
calorimetry. The same simulation including a
heavy ion background using the HIJING generator has been produced. The mean
reconstructed jet energy has then been corrected, on the average, looking at
the ratio of the reconstructed over generated jets as a function of the
reconstructed jet energy. This correction does not take into account
the smearing of the spectrum which is amplified from $p+p$
to $Pb+Pb$ collisions. Indeed, in a heavy ion UE
and due to the steeply falling shape of the input spectrum, even more
contributions at low $p_{T}$ populate the higher energetical part of
the reconstruted jet spectrum increasing its smearing. This of course
will have to be taken into account in a meaningful comparison of the
$N+N$ and $A+A$ data. In the present paper, an average correction has
been applied on the jet reconstructed energy so that the results
presented below on the HBP are still biased by the smearing effect.
\subsection{Background and quenching effects on the fragmentation function}
\begin{figure}[h]
\begin{center}
\includegraphics[scale=0.32, angle=0]{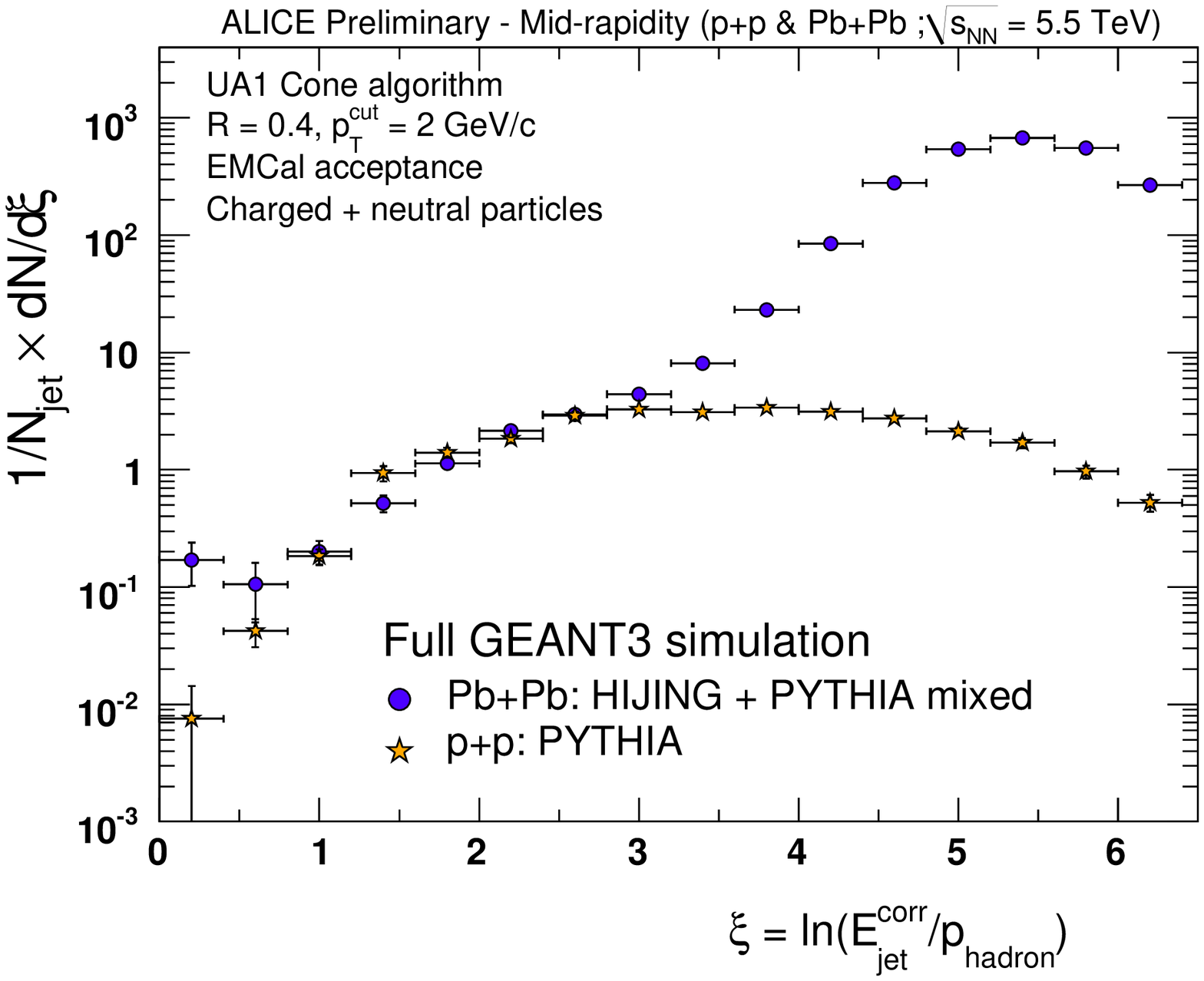}     
\includegraphics[scale=0.32, angle=0]{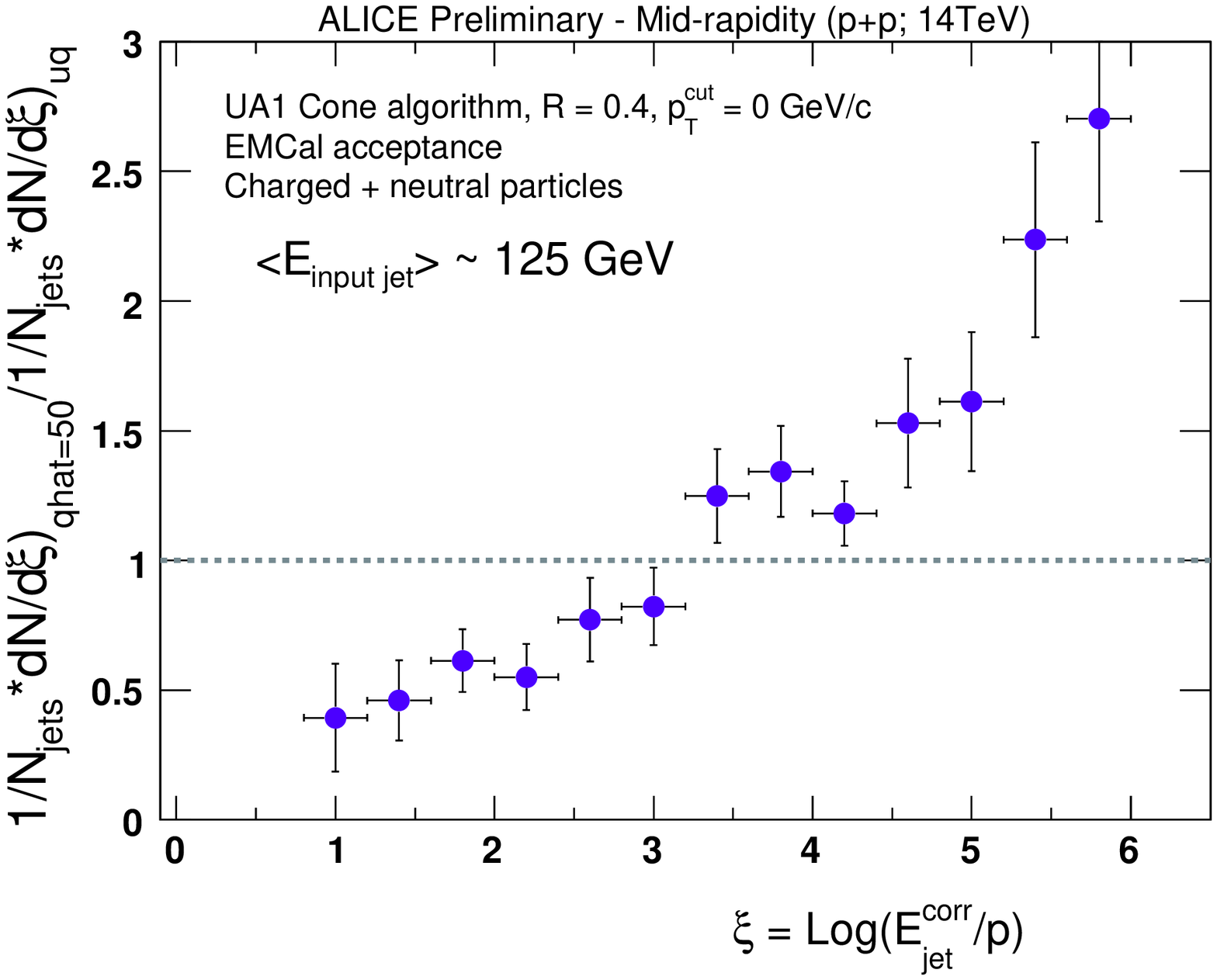}     
\caption{\underline{Left}: Hump-backed plateau in $p+p$ (stars) and $Pb+Pb$ collisions not background subtracted (circles) as a function of $\xi$ at $\sqrt{s_{NN}} = 5.5~$TeV. \underline{Right}: ratio of the HBP obtained in a $p+p$ quenched scenario over a non quenched one vs $\xi$ in $p+p$ collisions at $\sqrt{s_{NN}} = 14~$TeV.\label{fig:six}}
\end{center}
\end{figure}
Radiation phenomena in QCD and how they are
modified in a dense medium should be accurately probed
by understanding how the energy is distributed inside jets. Therefore,
it strongly motivates the study of the distribution of hadrons inside
jets: the HBP. Moreover, it offers a particular window of study on the
hadronisation phenomenon badly understood today.
It is
important to understand the effects of the heavy ion UE on its
extraction. The domain of interest of such distribution is for the
$\xi$ region dominated by the production of soft particles which come
from the gluon radiation emission in a quenching scenario. For jets of
energy $70-100~$GeV, this region turns out to be for a $\xi$ above
$\sim 3$. Figure~\ref{fig:six} (left) presents the modified
fragmentation function $1/N_{jet} \times dN/d\xi$ as a function of
$\xi = ln(E_{jet}^{Corr}/p_{hadron})$ in $p+p$ and $Pb+Pb$ collisions
at $\sqrt{s_{NN}} = 5.5~$TeV. The full jet spectra have been
considered here. In a first step, no quenching scenario has been
included in these simulations in order to understand how the soft
background of the UE by itself modifies the expected fragmentation
function. As seen in Fig.~\ref{fig:six}, the soft emission drastically
twists (more than 2 orders of magnitude) the HBP, increasing the
number of entries in the high $\xi$ region giving rise to a distortion
of the distribution. In order to go a step further in the comparison
of $p+p$ and $Pb+Pb$ HBP, the data have to be background
subtracted. Despite a good background subtraction, the data for $\xi >
5$ will not be exploitable anymore as dominated by too large error
bars. This background subtraction procedure and the results associated
are not presented here.

Instead, we have chosen to show the ratio of two HBP obtained in $p+p$
collisions at $\sqrt{s} = 14~$TeV with and without quenching
scenario to show the sensitivity one should expect vs $\xi$.
For such a distribution we assume a perfect
background subtraction procedure. Without specific trigger bias in the
data selection and for jets of $125~$GeV, one obtains a ratio which
increases with $\xi$ increasing with a value below one for a $\xi
\sim~3$ and above one after. Both amplitudes below and above this
$\xi$ limit, as well as the exact $\xi$ position of a ratio equals to
unity should allow us to quantify the strengh of the quenching scenario.
\section{Conclusion}
Technical aspects for jet reconstruction in $p+p$ and $A+A$ collisions
have been discussed.  More specifically, the expected performance for
jet physics studies in ALICE have been presented. The observation
of some modifications of the jet structure in $Pb+Pb$ collisions at LHC
will be possible for $\xi$ up to $\sim$ 5 where
we expect to see a clear distortion of the HBP due to
the soft emission generated by gluon radiation over the soft
background of the UE.

\begin{footnotesize}
\bibliographystyle{mpi08} 
{\raggedright
\bibliography{mpi08}
}
\end{footnotesize}
\end{document}